\documentclass[aps,prd,twocolumn,showpacs,superscriptaddress,amsmath,graphicx,psfrag,color,longtable]{revtex4}
\usepackage{psfrag}
\usepackage{color}
\usepackage{longtable}

\begin{document}

\title{On the flavor structure of the littlest Higgs model}

\author{Svjetlana Fajfer}
\email[Electronic address:]{svjetlana.fajfer@ijs.si}
\affiliation{J. Stefan Institute, Jamova 39, P. O. Box 3000, 1001 Ljubljana, Slovenia}
\affiliation{Department of Physics, University of Ljubljana, Jadranska 19, 1000 Ljubljana, Slovenia}

\author{Jernej F. Kamenik}
\email[Electronic address:]{jernej.kamenik@ijs.si}
\affiliation{J. Stefan Institute, Jamova 39, P. O. Box 3000, 1001 Ljubljana, Slovenia}
\affiliation{INFN, Laboratori Nazionali di Frascati, I-00044 Frascati, Italy}

\date{\today}

\begin{abstract}
We investigate the Yukawa sector for up-like quarks in the Lee's
version of the Littlest Higgs model. We derive general quark mass
and mixing formulae and study leading order contributions due to
non-zero light quark masses. Relying on the unitarity of the
generalized quark mixing matrix we obtain corrections to the CKM
matrix elements. In this model FCNCs appear at the tree level and
using leading order contributions we obtain the FCNC couplings for
the up-like quark transitions. In light of recent experimental
results on the $D^0 - \bar D^0$ transition we make predictions for
$x_D$ as well as the {$D \to \mu^+ \mu^-$} decay rate. Finally, we
discuss probabilities for the $t \to c (u) Z $ transitions relevant
for the LHC studies.
\end{abstract}


\maketitle

\section{Introduction}

The existence of the hierarchy problem within the SM stimulated
constructions  of many models of new physics. In the last three
decades supersymmetric models offered appealing solutions to the
hierarchy problem, although the existence of susyparticles has not
been confirmed experimentally. During the last few years, the Little
Higgs models~\cite{ArkaniHamed:2001nc,ArkaniHamed:2002pa,
ArkaniHamed:2002qx, ArkaniHamed:2002qy, Low:2002ws}  have attracted
a lot of attention offering an alternative solution to the hierarchy
problem. The main features of all Little Higgs-like  models are that
Higgs fields appear as Goldstone bosons of a global symmetry broken
at some new scale. Then they acquire masses and become
pseudo-Goldstone bosons via symmetry breaking at the electroweak
scale. The quadratic divergences in the Higgs mass due to the SM
gauge bosons are canceled by the contributions of the new heavy
gauge bosons with spin 1. The divergence due to the top quark is
canceled by the contribution of the new heavy vector-like quark with
the charge $2/3$ and spin $1/2$.

\par

In the simplest model {(named the Littlest Higgs model)} which has
been studied extensively in the literature
~\cite{Han:2003wu,Buras:2004kq,Buras:2006wk} the masses of $u$, $d$,
$s$, $c$ and $b$ quarks are usually neglected in comparison with the
electroweak symmetry breaking scale.
Consequently, some tree-level FCNCs appear in the up-quark sector,
but only coupling the new heavy quark to the top quark and the $Z$
boson. At the same time only $V_{tb}$ CKM matrix element receives
small corrections due to CKM non-unitarity. In a generalization of
that model given by Lee~\cite{Lee:2004me} mixing of the lighter
quarks with the top quark is present. There are two interesting
consequences that appear in such a scenario. It allows for
$Z$-mediated FCNCs at the tree-level in the whole up-quark sector
(while not in the down-quark sector). It also extends the $3\times
3$ CKM matrix in the SM to a $4\times 3$ matrix and introduces
non-unitarity corrections to all of the CKM matrix elements.
Recently, Chen et al.~\cite{Chen:2007yn} have discussed $D - \bar D$
mixing in a similar model, but only after imposing additional
assumptions. Namely, in order to preserve the large up-quark mass
hierarchy they assume a special form of the Yukawa matrices,
allowing them to constrain the model parameters. Using present
errors in the CKM matrix elements they are able to induce rather
large flavor changing effects.

\par

Motivated by the results of these papers we re-investigate the
flavor structure of the Littlest Higgs model (LHM).
We perform an eigensystem analysis of the more general LHM up-quark
mass matrix
and are able to recover the results of the
constrained model~\cite{Han:2003wu} as well as give robust
predictions for the more general case. After Introduction, we give a
general analysis of the up-quark Yukawa couplings in section II\,.
Section III contains analysis of CKM unitarity and FCNCs.
Phenomenological consequences are discussed in section IV, while
conclusions are given in section V\,.

\section{LHM up Yukawas and CP violation}

We first focus on the simplest LHM, whose phenomenology was first
studied by Han et al.~\cite{Han:2003wu}. The light and heavy top
quark Yukawa sector of this model is given by eq.~(24)
of~\cite{Han:2003wu}:
\begin{equation}
\mathcal L_{Y} = \frac{1}{2} \lambda_1 f \epsilon_{ijk} \epsilon_{xy} \chi_i \Sigma_{jx} \Sigma_{ky} u_3^{'c} + \lambda_2 f \tilde t \tilde t^{'c} + \mathrm{H.c.},
\label{eq:L_Y_Han}
\end{equation}
where $\chi^T = (b_3,t_3,\tilde t)$, $\epsilon_{ijk}$ and
$\epsilon_{xy}$ are antisymmetric tensors, with $ijk = 1,2,3$ and
$xy=4,5$. $\Sigma$ contains the Higgs fields in the adjoint
representation of the global LH $SU(5)$ (c.f.~\cite{Han:2003wu}
eq.~(3)), $u_3^{'c}$ and $\tilde t^{'c}$ are the two right-handed
top fields, while $f$ is the VEV of the heavy Higgs ($f\simeq
1~\mathrm{TeV}$). Note that $\lambda_{1,2}$ are $c-$numbers in this
model implying absence of mixing of the third generation with the
first two generations in the up sector. However, in this form, the
model is also CP conserving, as can be easily deduced by studying
weak basis invariants of the resulting mass
matrix~\cite{Aguila:1997vn,Branco:1999fs}. CP is preserved even in
presence of non-diagonal Yukawa terms involving only the first two
generations of up-quarks and regardless of the down-quark sector. In
order to provide SM-like sources of CP violation, one must therefore
add further non-diagonal Yukawa terms, mixing the light top quark
with the first two generations, but necessarily not involving the
heavy top quark. If we require that the one-loop top quark
contributions to the Higgs mass largely vanish, these additional
Yukawa couplings (we denote them by $\lambda^u_{ij}$) must be much
smaller than $\lambda_{1}$. This leads to a generalized up-quark
mass matrix in the weak basis
\begin{equation}
\mathcal M_p =
\begin{pmatrix}
 i v \lambda^u_{11} & i v \lambda^u_{12} & i v \lambda^u_{13} & 0 \\
 i v \lambda^u_{21} & i v \lambda^u_{22} & i v \lambda^u_{23} & 0 \\
 i v \lambda^u_{31} & i v \lambda^u_{32} & i v (\lambda_{1} + \lambda^u_{33}) & 0 \\
 0 & 0 & f \lambda_{1} & f \lambda_{2}
\end{pmatrix}.
\label{eq:M_p_Han}
\end{equation}

\par

%

Lee~\cite{Lee:2004me} similarly generalizes the Yukawa part of the
model by including a general mixing pattern in the up-quark sector.
In Eq.~(2.15) of~\cite{Lee:2004me} he writes
\begin{equation}
\mathcal L_{Y} = \frac{1}{2} \lambda_1^{ab} f \epsilon_{ijk} \epsilon_{xy} \chi_{a i} \Sigma_{jx} \Sigma_{ky} u_b^{'c} + \lambda_2 f \tilde t \tilde t^{'c} + \mathrm{H.c.},
\label{eq:L_Y_Lee}
\end{equation}
with $ab = 1,2,3$ and $\chi_i^T = (b_i,t_i,\delta_{i3} \tilde t)$.
Then the up-quark mass matrix in the weak basis should
become~\footnote{We find a difference in the fourth row calculation
of ref.~\cite{Lee:2004me}.}
\begin{equation}
\mathcal M_p =
\begin{pmatrix}
 i v \lambda_1^{11} & i v \lambda_1^{12} & i v \lambda_1^{13} & 0 \\
 i v \lambda_1^{21} & i v \lambda_1^{22} & i v \lambda_1^{23} & 0 \\
 i v \lambda_1^{31} & i v \lambda_1^{32} & i v \lambda_1^{33} & 0 \\
 f \lambda_1^{31} & f \lambda_1^{32} & f \lambda_1^{33} & f \lambda_{2}
\end{pmatrix}.
\label{eq:M_p_complete}
\end{equation}
We see that the mass matrices in the two models apparently differ in
the form of their fourth rows. However, when requiring (partial)
cancelation of top quark contributions to the Higgs mass, both
models can be treated equivalently.

\par

We perform an eigensystem analysis of this quark mass sector based
on the conjugated versions of eqs.~(24.26) of
ref.~\cite{Branco:1999fs}. Namely we can denote
\begin{equation}
\mathcal M_p^{\dagger} =
\begin{pmatrix}
 G_{p(3\times 3)} & J_{p (3 \times 1)} \\
 0 & \hat M_p
\end{pmatrix},
\label{eq:M_p_decomposition}
\end{equation}
while the down-quark mass matrix $M_n$ is general three-by-three and complex.
{$\hat M_p$ is a
 $c-$value 
and} can be made real via suitable phase redefinition of the heavy
top field $\tilde t^{'c}$, while $G_p$ can be made diagonal and real
via suitable weak basis transformations ($G_p\to \mathrm{diag( v
\eta_{1},v \eta_{2}, v \eta_{3})}$). The unitary transformations
involved induce corrections to $J_p$ in terms of mixing of
components which we since denote with tilde: $J_p^T=(f \tilde
\lambda_1^{31}, f \tilde \lambda_1^{32}, f \tilde \lambda_1^{33})$.
$\tilde \lambda_1^{3i} = \sum_j L_{ij} \lambda_1^{(3j)}$ with
$L_{ij}$ being components of a unitary matrix diagonalizing $G_p$ so
that $\sum_j |L_{ij}|^2 =1$ for any $i$. We see that the end form of
$\mathcal M_p$ is qualitatively the same for both models under
consideration.
The mass eigenvalue equation $\mathcal M_p \mathcal M_p^{\dagger} W_p = W_p D_p^2$, where
\begin{equation}
W_p=\begin{pmatrix}
 K_{p(3\times 3)} & R_{p (3 \times 1)} \\
 S_{p (1\times 3)} & T_p
\end{pmatrix}
\label{eq:W_p_decomposition}
\end{equation}
is a unitary eigenvector matrix and $D_p=\mathrm{diag}[\bar
m_{p(3\times 3)},\bar M_p]$ is the diagonal eigenmass matrix ($\bar
m_p=\mathrm{diag}(m_1,m_2,m_3)$), can then be written as a set of
matrix equations~\cite{Branco:1999fs}
\begin{subequations}
\begin{eqnarray}
\label{eq:eigenmass_a}G_p^{\dagger} G_p K_p + G_p^{\dagger} J_p S_p &=& K_p \bar m_p^2, \\
\label{eq:eigenmass_b}G_p^{\dagger} G_p R_p + G_p^{\dagger} J_p T_p &=& R_p \bar M_p^2, \\
\label{eq:eigenmass_c}J_p^{\dagger} G_p K_p + (J_p^{\dagger} J_p + M_p^2) S_p &=& S_p \bar m_p^2, \\
\label{eq:eigenmass_d}J_p^{\dagger} G_p R_p + (J_p^{\dagger} J_p + M_p^2) T_p &=& T_p \bar M_p^2,
\end{eqnarray}
\end{subequations}
while the $W_p$ unitary constraint relevant for this discussion reads
\begin{equation}
R_p^{\dagger} R_p + T_p^* T_P = 1.
\end{equation}

\par

We start by evaluating eqns.~(\ref{eq:eigenmass_b} and \ref{eq:eigenmass_d}):
\begin{subequations}
\begin{eqnarray}
R_p \bar M_p^2 &=& v^2 \mathrm{diag}(\eta_{1}^2,\eta_{2}^2,\eta_{3}^2)  R_p \nonumber\\
&&+ v f (\eta_{1} \tilde \lambda_1^{31},\eta_{2} \tilde \lambda_1^{32},\eta_{3} \tilde  \lambda_1^{33})^T T_p\,,   \\
T_p \bar M_p^2  &=& v f (\eta_1 \tilde \lambda_1^{31*},\eta_2 \tilde \lambda_1^{32*},\eta_3 \tilde  \lambda_1^{33*}) R_p \nonumber\\
&&+ f^2 |\lambda|^2 T_p,
\end{eqnarray}
\end{subequations}
where $|\lambda|^2 = (|\tilde\lambda_1^{31}|^2+|\tilde\lambda_1^{32}|^2+|\tilde\lambda_1^{33}|^2+|\lambda_2|^2)$. We notice that requiring the heavy top mass to scale as $\bar M_p \sim f$ the two equations can be solved simultaneously provided $R_p \lesssim T_p$ in terms of $v/f$ scaling. Then, to leading order in $v/f$, the heavy top mass is
\begin{equation}
\bar M_p^2 = |\lambda|^2 f^2,
\end{equation}
while for $R_p$ and $T_p$ we get
\begin{subequations}
\begin{eqnarray}
\label{eq:R_p_sol}R_p &=& \frac{v}{f} \frac{1}{|\lambda|^2} (\eta_{1} \tilde \lambda_1^{31},\eta_{2} \tilde \lambda_1^{32},\eta_{3} \tilde  \lambda_1^{33})^T T_p, \\
|T_p| &\simeq& 1- \frac{1}{2}R_p^{\dagger} R_p = 1-\mathcal O(v/f)^2\,,
\end{eqnarray}
\end{subequations}
where the unitarity constraint together with $v/f$ expansion of the
square root has been used in the last line.

\par

Next we evaluate eqns.~(\ref{eq:eigenmass_a} and \ref{eq:eigenmass_c})
\begin{subequations}
\begin{eqnarray}
\label{eq:eigenmass_factor_a}K_p \bar m_p^2 &=& v^2 \mathrm{diag}(\eta_{1}^2,\eta_{2}^2,\eta_{3}^2)  K_p \nonumber\\
&&+ v f (\eta_{1} \tilde \lambda_1^{31},\eta_{2} \tilde \lambda_1^{32},\eta_{3} \tilde  \lambda_1^{33})^T S_p\,,   \\
\label{eq:eigenmass_factor_c}S_p \bar m_p^2  &=& v f (\eta_1 \tilde \lambda_1^{31*},\eta_2 \tilde \lambda_1^{32*},\eta_3 \tilde  \lambda_1^{33*}) K_p \nonumber\\
&&+ f^2 |\lambda|^2 S_p\,.
\end{eqnarray}
\end{subequations}
Requiring the light up-quark mass eigenvalues to scale as $\bar m_p
\sim v$, we can solve both equations without any fine-tuning
provided $S_p$ and $K_p$ have fixed relative scaling in $v/f$: $S_p
\sim K_p v/f$. Then the left hand side of
eq.~(\ref{eq:eigenmass_factor_c}) is of higher order in $v/f$ than
the right hand side and can be neglected yielding the relation
\begin{equation}
S_p = -\frac{v}{f} \frac{1}{|\lambda|^2} (\eta_{1} \tilde \lambda_1^{31*},\eta_{2} \tilde \lambda_1^{32*},\eta_{3} \tilde  \lambda_1^{33*}) K_p\,.
\end{equation}
Inserting this expression into eq.~(\ref{eq:eigenmass_factor_a})
yields~\footnote{The appearance of the off-diagonal contributions in
eq.~(\ref{eq:eignemass_factor_a2}) is the direct consequence and
main difference due the different $v/f$ scaling of $J_p$ with
respect to the one in section 24.3 of ref.~\cite{Branco:1999fs}.}
\begin{eqnarray}
K_p \mathrm{diag}(m_1^2,m_2^2,m_3^2) &=& v^2 \left[ \mathrm{diag}(\eta_{1}^2, \eta_{2}^2, \eta_{3}^2 ) \right.\nonumber\\
&&- \left.\left (\frac{\eta_{i}\eta_{j} \tilde \lambda_1^{3i}\tilde\lambda_1^{3j*} }{|\lambda|^2}\right)_{(3 \times 3)} \right] K_p\,, \nonumber\\
\label{eq:eignemass_factor_a2}
\end{eqnarray}
where in this short-hand matrix notation there is {\it no} summation
over the repeated quark generation indices. The above matrix
equation in full form is given in the appendix. Next we notice that
the off-diagonal elements of the matrix multiplying $K_p$ on the
right hand side of eq.~(\ref{eq:eignemass_factor_a2}) are generally
smaller than the diagonal ones and tend to zero with $\tilde
\lambda_1^{i3} / \lambda_2 \to 0$. Therefore we approximate the
solution, unitary at leading order in $v/f$, with a linear expansion
around the diagonal, yielding
\begin{subequations}
\begin{eqnarray}
\label{eq:eigenmass_solutions_a}
m_i^2 &=& v^2 \eta_{i}^2 \left[ 1 - \frac{|\tilde \lambda_1^{3i}|^2}{ |\lambda|^2} \right], \\
(K_{p})_{ij} &=& \delta_{ij} + (\delta_{ij}-1) \frac{v^2 \eta_{i} \eta_{j} \tilde \lambda_1^{3i}\tilde \lambda_1^{3j*}}{(m_i^2-m_j^2)|\lambda|^2}\,.\nonumber\\
\label{eq:eigenmass_solutions_b}
\end{eqnarray}
\end{subequations}
Again in eq.~(16b) there is no summation over repeated quark
generation indices and the full matrix form of $K_p$ in this
approximation is given in the appendix. With $\tilde
\lambda_1^{31}=\tilde \lambda_1^{32}=0$ and $\tilde
\lambda_1^{33}=\eta_3 = \lambda_{1}$ we reproduce the usual result
for the light and heavy top masses in the simplest model{ of Han et
al.}~\cite{Han:2003wu} which ensures exact cancelation of top-quark
contributions to the Higgs mass at one loop. Deviations from this
limit in terms of non-vanishing $\tilde \lambda_1^{31}$ and $\tilde
\lambda_1^{32}$ on one side reintroduce such corrections, while on
the other side they provide needed sources of SM-like CP violation.

\section{CKM unitarity and FCNC{s}}

FCNCs at tree level via flavor changing Z couplings can be easily
deduced by evaluating $Z_p = A_p^{\dagger} A_p$, where $A_p$ are the
first three rows of $W_p$ or $A_p=(K_p,R_p)$. Then the FCNC of
up-like quarks coupling to the $Z$ boson is $J^{FC}_{\mu}= (g/2 c_W)
\bar u_{Li} \gamma_{\mu} (Z_p)_{ij} u_{Lj}$, where $g$ is the
$SU(2)_L$ gauge coupling and $c_W$ is the cosine of the Weinberg
angle. At leading order in $v/f$ we get off-diagonal elements of
$Z_p$ only in the fourth column (and row)
\begin{eqnarray}
(Z_p)_{i4} &=& \sum_{j} (K_p)^*_{ji} (R_{p})_j
\label{eq:Z_pi4}
\end{eqnarray}
FCNCs among the light up-type quarks only come at the order of
$\mathcal(v/f)^2$, are due to $4 \times 4$ up-quark basis
unitarity~\cite{Branco:1999fs} and yield
\begin{eqnarray}
(Z_p)_{ij} &=& \delta_{ij}-(Z_p)^*_{i4}(Z_p)_{j4}.
\label{eq:Z_pij}
\end{eqnarray}

\par

At the same time we get CKM non-unitary corrections in terms of
fourth row CKM matrix elements, which can be calculated via $V_{CKM}
= A_p^{\dagger} A_n$, where $A_n$ is the $3 \times 3$ down quark
unitary mixing matrix. In absence of fourth row entries in $\mathcal
M_p$ due to the mixing with the vector top quark, $A_p$ would just
be the identity and the usual form of $V_{CKM}=A_{n}$ would be
obtained. Now however, we obtain for the fourth row CKM matrix
elements
\begin{equation}
(V_{CKM})_{4i} = \sum_k (R_p)^*_k (A_n)_{ki}\,,
\label{eq:V_CKM4i}
\end{equation}
while the $3\times 3$ non-unitary mixing submatrix for the light
quarks is, again due to $4\times 4$ unitarity
\begin{equation}
(V_{CKM})_{ij} = \sum_k (K_p)^*_{ki} (A_n)_{kj} - (V_{CKM})^*_{4i}(V_{CKM})_{4j}\,.
\label{eq:V_CKMij}
\end{equation}
Formulae~(\ref{eq:Z_pi4}-\ref{eq:V_CKMij}) are exact up to $v/f$
corrections, but more importantly {\it regardless of any
approximations to the solution for $K_p$ from
eq.~(\ref{eq:eignemass_factor_a2})}, thus representing faithfully
the generally rich flavor structure of the LH model.

\par

Our treatment leads to qualitatively similar conclusions as found
in~\cite{Lee:2004me} regarding FCNCs, but we disagree in the
procedure as well as in the form of the final results. The approach
of~\cite{Chen:2007yn} on the other hand imposes fine-tuning
cancelations among up-quark Yukawa elements (i.e. requiring
cancelation of the two terms in the square brackets in (16a) for the
first two generations) in order to obtain relations among them.
However not all parameters feature in independently in the mass
formulae. By identifying the heavy top mass  $m_T=f
\sqrt{|\lambda^2|}$, we find that all expressions only depend on
certain combinations: $(v \eta_{i})$ and $e_i \equiv \tilde
\lambda_1^{3i} /\sqrt{|\lambda^2|}$. Using the first, we can absorb
all light Higgs VEV dependence into light quark masses and mixings,
while the second indicates that phenomenologically, the LH FCNC
couplings lie on three-plane intersection of a four-sphere with
radius $\sqrt{|\lambda^2|}$. Therefore we parameterize the moduli of
$e_i$ using generalized Euler's angles, projected on the three-plane
(distance from the origin is parameterized by $\sin\gamma$)
$\alpha,\beta,\gamma$: $|e_1| = |s_{\alpha}s_{\beta}s_{\gamma}|$,
$|e_2| = |c_{\alpha}s_{\beta}s_{\gamma}|$, $|e_3| =
|c_{\beta}s_{\gamma}|$, where $s_x = \sin x$ and $c_x = \cos x$.
Note that, although $|e_i|$ are bounded to lie between 0 and 1,
providing sources of SM like CP violation discussed in the previous
section requires at least two of them to be different from zero (the
constrained model of Han corresponds to $c_{\beta}=1$ or
$e_1=e_2=0$). At the same time, due to the orthogonality of
projections $c_x$ and $s_x$, only one of the $|e_i|$ can be set
close to 1 at best, while in addition cancelation of top loop
contributions to the Higgs mass requires $|e_3|$ to be much larger
than $|e_{1,2}|$. This eventually rules out a simultaneous mass
cancelation via fine-tuning for the first two generations in
eq.~(16a).

\par

More explicit analytic expressions for CKM corrections and FCNCs in
closed form can then be obtained by keeping only the leading order
terms in the off-diagonal expansion of $K_p$ (i.e. using
solutions~(\ref{eq:eigenmass_solutions_a})
and~(\ref{eq:eigenmass_solutions_b})) in which case our analysis
reverts to the one of ref.~\cite{Branco:1999fs}. We obtain
\begin{subequations}
\begin{eqnarray}
(Z_p)_{i4} &\simeq& \frac{m_i}{m_T} \frac{e_i}{ \sqrt{1-|e_i|^2}}\simeq (R_p)_i\,.\\
(Z_p)_{ij} &\simeq& \delta_{ij}-\frac{m_i m_j}{m_T^2} \frac{e^*_i}{\sqrt{1-|e_i|^2}} \frac{e_j}{\sqrt{1-|e_j|^2}} \,,\nonumber\\
\\
(V_{CKM})_{4i} &\simeq& \sum_k \frac{m_k}{m_T} (A_n)_{ki} \frac{e^*_k}{\sqrt{1-|e_k|^2}} \nonumber\\
&\simeq& \sum_k \frac{m_k}{m_T} (V_{CKM})_{ki} \frac{e_k}{\sqrt{1-|e_k|^2}}\,,\\
(V_{CKM})_{ij} &\simeq& (A_n)_{ij} - (V_{CKM})^*_{4i}(V_{CKM})_{4j}\,.
\end{eqnarray}
\end{subequations}
Actually, due to the large hierarchy in the up quark masses,
expansion~(16b) is always a good approximation for $K_p$. This can
be seen by parameterizing the off-diagonal elements of $K_p$ in
eq.~(16b) or~(\ref{eq:Kp_App}) in terms of generalized Euler's
angles and physical quark masses. Then due to the orthogonality of
the projections $c_i$, $s_i$ expressions of the type $e^*_i e_j /
\sqrt{1-|e_i|^2}\sqrt{1-|e_j|^2}$ for $i\neq j$ are always bounded
from above by 1, while off-diagonal elements in $K_p$ are in
addition suppressed by small ratios of quark masses among different
generations. Therefore, even if the eigenvalues in eq.~(16a) receive
relatively large corrections, these are not reflected in large
deviations from the diagonality in $K_p$ and consequently in FCNCs
as we will see in the next section.

\section{Phenomenology}

%
%
%

\begin{figure}
\scalebox{0.8}{\includegraphics{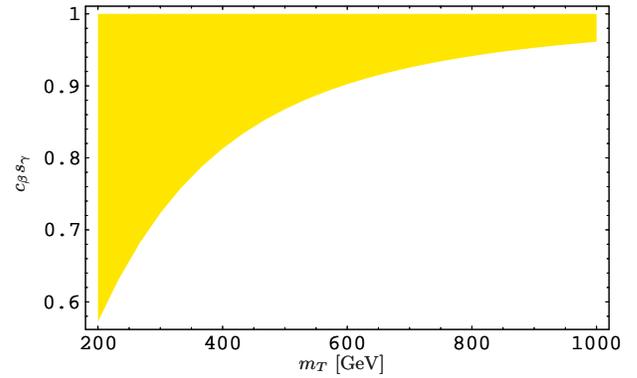}}
\caption{\label{fig:Zp33}LHM parameter plane spanned by $m_T$ and
$c_{\beta} s_{\gamma}$. The shaded region in yellow(grey) is
excluded by present CKM unitarity bounds as explained in the text.}
\end{figure}

We first calculate FCNC constraints, given experimentally from CKM
non-unitarity~\cite{Lee:2004me}. We take the current bounds on the
CKM moduli, obtained from tree level processes without referring to
$3\times 3$ CKM unitarity~\cite{PDBook}. Then the complete $4\times
4$ mixing matrix unitarity conditions constrain FCNCs through the
relation $Z_p=V_{CKM} V_{CKM}^{\dagger}$~\cite{Branco:1999fs}. We
notice that the stringiest unitarity bounds on the parameters will
come from the top sector due to large up-quark mass hierarchy, and
from the diagonal elements, where the couplings are not bounded by
orthogonality conditions. In particular, the most constraining is
the recent direct lower bound on the magnitude of the $V_{tb}$ CKM
matrix element $|V_{tb}|>0.78$ from the D0
collaboration~\cite{Abazov:2006bh}. We write down the most
perspective constraints
\begin{subequations}
\begin{eqnarray}
\label{eq:Zp33}
|(Z_p)_{33}| &=& \left| 1-\frac{m_t^2}{m_T^2} \frac{c^2_{\beta} s^2_{\gamma}}{(1-c^2_{\beta} s^2_{\gamma})} \right| > 0.63\,, \\
|(Z_p)_{32}| &=& \left| \frac{m_t m_c}{m_T^2} \frac{s_{\alpha} c_{\beta} s_{\beta} s^2_{\gamma}}{\sqrt{1-s^2_{\alpha} s^2_{\beta} s^2_{\gamma}}\sqrt{1-c^2_{\beta} s^2_{\gamma}}} \right| < 0.13\,, \nonumber\\
&&\\
|(Z_p)_{31}| &=& \left| \frac{m_t m_u}{m_T^2} \frac{c_{\alpha} c_{\beta} s_{\beta} s^2_{\gamma}}{\sqrt{1-c^2_{\alpha} s^2_{\beta} s^2_{\gamma}}\sqrt{1-c^2_{\beta}  s^2_{\gamma}}} \right| < 0.11\,, \nonumber\\
&&\\
|(Z_p)_{22}| &=& \left| 1-\frac{m_c^2}{m_T^2} \frac{s^2_{\alpha} s^2_{\beta} s^2_{\gamma}}{1-s^2_{\alpha} s^2_{\beta} s^2_{\gamma}} \right| > 0.63\,, \\
|(Z_p)_{11}| &=& \left| 1-\frac{m_u^2}{m_T^2} \frac{c^2_{\alpha}
s^2_{\beta} s^2_{\gamma}}{1-c^2_{\alpha} s^2_{\beta} s^2_{\gamma}}
\right| > 0.97\,.
\end{eqnarray}
\end{subequations}
Presently only eq.~(\ref{eq:Zp33}) is restrictive enough to be used
as any kind of constraint on the parameters of the model. It
excludes a region in the parameter plane spanned by $m_T$ and
$c_{\beta} s_{\gamma}$ (corresponding to
$\lambda_1/\sqrt{\lambda_1^2+\lambda_2^2}$ in the constrained model)
as shown on fig.~\ref{fig:Zp33}. \psfrag{pijl}[bc]{$\pi^i(q)$}
\psfrag{x}[bc]{$m_T~[\mathrm{GeV}]$} \psfrag{y}[bc]{$c_{\beta}
s_{\gamma}$}
We see that for heavy top-quark masses above $1$~TeV, even this
bound is ineffective at present.

\par

Next we study $D-\bar D$ mixing. For $x_D=\Delta m_D/\Gamma_D$
contribution due to $Z$ mediated FCNCs we use the known form
\begin{equation}
x_D = \frac{\sqrt 2 m_D}{3 \Gamma_D} G_F f^2_D B_D |(Z_p)_{12}|^2 r_1(m_c,m_Z),
\end{equation}
where the function $r_1(\mu,M)=[\alpha_s(M)/\alpha_s(m_b)]^{6/23}$
$\times[\alpha_s(m_b)/\alpha_s(\mu)]^{6/25}$ accounts for the
one-loop QCD running, $G_F$ is the Fermi constant, $f_D$ is the $D$
meson decay constant and $B_D$ is the $D$ meson bag parameter. In
our numerical evaluation we use PDG~\cite{PDBook} values for quark
and $Z$ boson masses, mass and width of the $D$ meson, $G_F$ and
$\alpha_s(m_Z)$, while for the hadronic parameters we take
$f_D=0.22~\mathrm{GeV}$~\cite{Artuso:2005ym} from CLEO-c measurement
and $B_D=0.82$~\cite{Gupta:1996yt} from a quenched lattice study.
After evaluating these known quantities we obtain
\begin{eqnarray}
&&x_D = 2 \times 10^5 |(Z_p)_{12}|^2 \nonumber\\
&&\simeq 3 \times 10^{-12} \left|\frac{s_{\alpha} c_{\alpha} s^2_{\beta} s^2_{\gamma}}{\sqrt{1-c^2_{\alpha} s^2_{\beta} s^2_{\gamma}}\sqrt{1-s^2_{\alpha} s^2_{\beta} s^2_{\gamma}}} \left(\frac{1~\mathrm{TeV}}{m_T}\right)^2 \right|^2.\nonumber\\
\label{eq:DDbar}
\end{eqnarray}
We have to compare this expression with the recent experimental
results from the $B$-factories~\cite{Staric:2007dt,Aubert:2007wf},
which give a value of $x_D=0.0087\pm0.003$~\cite{Fajfer:2007dy}.
Similarly for the rare $D\to \mu^+ \mu^-$ decay width, we use the
known form for $Z$ mediated FCNC contribution
\begin{equation}
\Gamma(D^0\to \mu^+\mu^-) = \frac{m_D}{64\pi}\left(\frac{G_F}{\sqrt 2}\right)^2 |(Z_p)_{12}|^2 f_D^2 m_{\mu}^2\sqrt{1-\frac{4 m_{\mu}^2}{m_D^2}}
\end{equation}
and obtain
\begin{eqnarray}
&&\mathcal Br(D^0\to \mu^+ \mu^-) = 3\times 10^{-4}|(Z_p)_{12}|^2 \nonumber\\
&&\simeq 3\times 10^{-21} \left|\frac{s_{\alpha} c_{\alpha} s^2_{\beta} s^2_{\gamma}}{\sqrt{1-c^2_{\alpha} s^2_{\beta} s^2_{\gamma}}\sqrt{1-s^2_{\alpha} s^2_{\beta} s^2_{\gamma}}} \left(\frac{1~\mathrm{TeV}}{m_T}\right)^2 \right|^2,\nonumber\\
\label{eq:Dmumu}
\end{eqnarray}
again to be compared to the current experimental limit $\mathcal
Br(D^0\to \ell^+ \ell^-)<1.2 \times 10^{-6}$~\cite{Aubert:2004bs}
from BaBar. We see that in both processes, the LHM contributions at
tree level are negligible. Note however, that due to the same FC $Z$
coupling appearing in both eqs.~(\ref{eq:DDbar})
and~(\ref{eq:Dmumu}) a general upper bound prediction for the
$Br(D^0\to \mu^+ \mu^-)$ mediated by such effective couplings can be
made. Namely, saturating the measured value of $x_D$ with the short
distance contribution in the first line of eq.~(\ref{eq:DDbar}) we
obtain an upper bound on $|(Z_p)_{12}| < 2 \times 10^{-4}$ and
consequently $Br(D^0\to \mu^+ \mu^-)_{Z_p}<2\times 10^{-11}$. The
rare $D$ decays due to $c \to u Z$ transitions are then also very
suppressed as already noticed in~\cite{Fajfer:2005ke,
Fajfer:2007dy}. Therefore we only give predictions for the $t\to c
Z$ and $t\to u Z$ decay rates. In the SM these transitions are
highly suppressed and their branching ratios are of the order
$\mathcal O(10^{-10})$ or less~\cite{AguilarSaavedra:2002ns}. On the
other hand, current experimental constraints on these transitions
are not very strong~\cite{PhysRevLett.82.1628}.
Following~\cite{AguilarSaavedra:2002ns,Lee:2004me}, we normalize the
decay width
\begin{equation}
\Gamma(t \to c (u) Z) = \frac{m_t^3}{16\pi}\frac{G_F}{\sqrt 2} |(Z_p)_{32(1)}|^2 f(x_Z,x_c),
\end{equation}
where $f(x,y)=[ (1-y)^2 - 2 x^2 + x (1+y) ] \lambda^{1/2}(x,y)$,
$\lambda^{1/2}(x,y) = \sqrt{1+y^2+x^2-2xy-2x-2y}$ and $x_i =
m_i^2/m_t^2$, to the dominant $t \to b W$ decay
rate~\cite{Donoghue:1992dd,AguilarSaavedra:2002ns}
\begin{equation}
\Gamma(t \to b W) = \frac{m_t^3}{8\pi}\frac{G_F}{\sqrt 2} |V_{tb}|^2 f(x_W,x_b),
\end{equation}
and obtain for the branching ratios approximately
\begin{eqnarray}
&&\mathcal Br(t\to c Z) \lesssim 0.5 \left|\frac{(Z_p)_{32}}{V_{tb}}\right|^2 \nonumber\\
&&\simeq 4\times 10^{-8} \left|\frac{s_{\alpha} c_{\beta} s_{\beta} s^2_{\gamma}}{\sqrt{1-c^2_{\beta} s^2_{\gamma}}\sqrt{1-s^2_{\alpha} s^2_{\beta} s^2_{\gamma}}} \left(\frac{1~\mathrm{TeV}}{m_T}\right)^2 \right|^2,\nonumber\\
\label{eq:tcZ}
\end{eqnarray}
and
\begin{eqnarray}
&&\mathcal Br(t\to u Z) \lesssim 0.5 \left|\frac{(Z_p)_{31}}{V_{tb}}\right|^2 \nonumber\\
&&\simeq 2\times 10^{-13} \left|\frac{c_{\alpha} c_{\beta} s_{\beta} s^2_{\gamma}}{\sqrt{1-c^2_{\alpha} s^2_{\beta} s^2_{\gamma}}\sqrt{1-c^2_{\beta}  s^2_{\gamma}}} \left(\frac{1~\mathrm{TeV}}{m_T}\right)^2 \right|^2,\nonumber\\
\label{eq:tuZ}
\end{eqnarray}
where in the last lines of eqs.~(\ref{eq:tcZ}) and~(\ref{eq:tuZ}) we
have again used the lower bound on $|V_{tb}|$
from~\cite{Abazov:2006bh}.

\section{Conclusions}

We have reinvestigated the LH model of Lee~\cite{Lee:2004me} by
applying general constraints on extra vector-like quark singlet
models given in ref.~\cite{Branco:1999fs}. Namely, we have discussed
the appearance of tree level FCNCs and CKM unitarity violation in a
LHM with general Yukawa couplings and shown that, contrary to
previous conclusions, the up-quark flavor changing $Z$ couplings are
{\it not} proportional to the CKM matrix elements. Instead they are
proportional to ratios of up-quark masses relative to the heavy top
quark mass and can be parameterized in terms of three new angle
parameters. Due to the large constraints on the heavy top quark
mass, these tree level contributions are found to be negligible even
when compared to SM loop contributions. Contrary to the derivation
of Chen et al.~\cite{Chen:2007yn}, we do not impose any fine tuning
and cancelations among the various Yukawa matrix elements in order
to obtain the measured up-quark masses. On the other hand, our
analysis shows, that mass relation between the light and heavy top
quark, ensuring the exact cancelation of one-loop contributions to
the Higgs mass, is not maintained in the general model. Relaxing
this requirement could have important effects on the currently
established heavy top quark mass limits from low energy
phenomenology.

 \begin{acknowledgments}
 This work is supported in part by the European Commission RTN network,
Contract No. MRTN-CT-2006-035482 (FLAVIAnet) and by the Slovenian
Research Agency.
 \end{acknowledgments}

\appendix

\section{Full forms of quark mass diagonalization formulae}

Here we give the matrix formulae given in short-hand notation in
eqs.~(15) and~(16b) in their full form and by using the
parametrization in terms of $e_{i}=\tilde \lambda_1^{3i}/|\lambda|$
parameters. Matrix eq.~(15) for $K_p$ reads
\begin{eqnarray}
&&K_p .
\begin{pmatrix}
m_1^2 & 0 & 0 \\
0 & m_2^2 & 0 \\
0 & 0 & m_3^2 \\
\end{pmatrix} \nonumber\\
&&= v^2
\begin{pmatrix}
    \eta_1^2 (1 - |e_1|^2) & - \eta_1 \eta_2 e_1 e_2^{*} & - \eta_1 \eta_3 e_1 e_3^{*}\\
     - \eta_1 \eta_2 e_2 e_1^{*} & \eta_2^2 (1 - |e_2|^2) & - \eta_2 \eta_3 e_2 e_3^{*}\\
     - \eta_1 \eta_3 e_3 e_1^{*} & - \eta_2 \eta_3 e_3 e_2^{*} & \eta_3^2 (1 - |e_3|^2)\\
\end{pmatrix}
.K_p\,,\nonumber\\
\end{eqnarray}
with the approximate solutions for $K_p$ of the form
\begin{equation}
K_p =
\begin{pmatrix}
    1 & - \frac{v^2 \eta_1 \eta_2 e_1 e_2^*}{(m_1^2-m_2^2)} & - \frac{v^2 \eta_1 \eta_3 e_1 e_3^*}{(m_1^2-m_3^2)}\\
    - \frac{v^2 \eta_1 \eta_2 e_2 e_1^*}{(m_2^2-m_1^2)} &1 &  - \frac{v^2 \eta_2 \eta_3 e_2 e_3^*}{(m_2^2-m_3^2)}\\
    - \frac{v^2 \eta_1 \eta_2 e_3 e_1^*}{(m_3^2-m_1^2)} & - \frac{v^2 \eta_2 \eta_3 e_3 e_2^*}{(m_3^2-m_2^2)} & 1\\
\end{pmatrix}.
\end{equation}
We remaind the reader that in this approximation the light up-quark
masses are given by $m_i = v \eta_i \sqrt{1-|e_i|^2}$. Then due to
the large measured mass hierarchy in the up-quark sector we have
approximately
\begin{equation}
\label{eq:Kp_App}
K_p \simeq
\begin{pmatrix}
    1 & \frac{m_1}{m_2} \hat e_1 \hat e_2^* & \frac{m_1}{m_3} \hat e_1 \hat e_3^*\\
    - \frac{m_1}{m_2} \hat e_2 \hat e_1^* & 1 & \frac{m_2}{m_3} \hat e_2 \hat e_3^*\\
    - \frac{m_1}{m_3} \hat e_3 \hat e_1^* & -\frac{m_2}{m_3} \hat e_3 \hat e_2^* & 1\\
\end{pmatrix},
\end{equation}
where we have used $\hat e_i = e_i/\sqrt{1-|e_i|^2}$.
\bibliography{article}

\end{document}